\documentclass[%
 reprint,
 amsmath,amssymb,
 aps,
]{revtex4-2}
\usepackage{amsmath,amssymb,bm}
\usepackage{graphicx,graphics,color}
\usepackage[colorlinks=true,linkcolor=green,bookmarks=true]{hyperref}
\usepackage{epsfig}
\usepackage{latexsym}
\usepackage{rotating}
\usepackage{subfigure}
\usepackage{centernot}
\usepackage{longtable}
\usepackage{hhline,multirow,tabularx}  % for nicer tables
\usepackage[utf8]{inputenc}
\usepackage{newunicodechar}
\newunicodechar{，}{,}
\usepackage{float}
\usepackage{mathrsfs}
\usepackage{xr}
\usepackage{adjustbox}
\usepackage{verbatim}

\newcommand{\RNum}[1]{\uppercase\expandafter{\romannumeral #1\relax}}

\usepackage{lipsum}

\usepackage{graphicx}% Include figure files
\usepackage{dcolumn}% Align table columns on decimal point
\usepackage{bm}% bold math
\usepackage{appendix}
\usepackage{microtype}

\usepackage{graphicx}% Include figure files
\usepackage{dcolumn}% Align table columns on decimal point
\usepackage{bm}% bold math
\begin{document}

\title{Nucleon  electromagnetic form factors in   nonlocal chiral  effective theory}

\author{Y. Salamu}
\author{Alim Ablat}
\affiliation{School of Physics and Electrical Engineering, Kashi University, Kashgar, 844000, Xinjiang, China}
\begin{abstract}
 In this paper, we calculate the pion one-loop corrections to the electromagnetic form factors of the nucleon up to $\mathcal{O}(p^4)$ order in the nonlocal framework. To systematically investigate the role of vector-meson dominance in the model, we also include the leading-order nucleon-vector-meson coupling. Owing to the renormalization-scheme dependence of chiral low-energy coupling constants (LECs), we refit these constants to the  nucleon magnetic moments, electromagnetic radii, and the $Q^2$-dependence of the electromagnetic form factors. Compared with the previous local analysis, the model results show that, in the nonlocal framework, the $Q^2$-dependence of the electromagnetic form factors in the large-$Q^2$ region is  significantly improved and are consistent  with experimental data.  
\end{abstract}

\date{\today}
\maketitle
%%%%%%%%%%%%%%%%%%%%%%%%%%%%%%%%%%%%%%%%%%%%%%%%%%%%%%%%%%%%%%%%%%%%%%%%
\section{Introduction}
The nucleon electromagnetic form factors are important observables in electron-nucleon elastic scattering and encode the fundamental electromagnetic structure of the nucleon. Over the past seven decades, as an important task in high-energy experiments, the extraction of nucleon electromagnetic form factors from electron-nucleon elastic scattering has made significant progress and revealed abundant data. Recently, attention has  been drawn to the Lattice QCD simulations of the electromagnetic form factors of the nucleon \cite{Djukanovic:2023beb,Alexandrou:2018sjm,Park:2021ypf}. This makes it possible to study the nucleon electromagnetic structure from the theoretical perspective. On the other hand, at the low energy scale, the interaction between low-lying hadrons is  described by  chiral perturbation theory (CHPT). As an important application, the electromagnetic properties of the nucleon have been investigated in chiral perturbation theory with various renormalization approaches, such as heavy-baryon (HB), infrared(IR) and extended-on-mass-shell(EOMS) regularization schemes \cite{Gasser:1987rb,Bernard:1992qa,Bernard:1998gv,Fearing:1997dp}. Although these results successfully explained the chiral expansion of magnetic moments, electromagnetic radii and $Q^2$-dependence of the electromagnetic form factors in the small-$Q^2$ region, the discrepancy between experimental data and model results in the large-$Q^2$ region is not yet well understood, even when the vector meson contributions are included \cite{Kubis:2000zd,Fuchs:2003ir,Schindler:2005ke}. 

As a novel approach for   renormalizing  the loop contributions, the nonlocal (or finite-range) renormalization method has been successfully applied in gravity and other areas of phenomenological studies \cite{Efimov:1967pjn,Alebastrov:1972dxn,Tomboulis:2015gfa,Kleppe:1991rv,Boos:2020ccj,Mazumdar:2018xjz,Ohta:1990fi,Terning:1991yt,Anikin:1995cf,Biswas:2014tua}. The main advantage of the nonlocal renormalization method is that the renormalization procedure is encoded into a modified nonlocal Lagrangian. Consequently, loop contributions from such a Lagrangian are UV convergent \cite{Donoghue:1998bs,Leinweber:2003dg,Young:2002ib}. More importantly, the nonlocal Lagrangians still fulfill the underlying symmetries of the theory with the help of the corresponding gauge-link operators \cite{Faessler:2006ft,Ivanov:1996fj}. This feature is reflected in the conservation of currents, which holds even at the loop level. Recently, the nonlocal renormalization method has been extended to  chiral perturbation theory, and within this model, pion-nucleon splitting functions as well as the nucleon electromagnetic form factors have been calculated \cite{He1,He2,He4,Wang03,He3,Salamu1,Salamu2}. However, we note that in the previous nonlocal analysis, the higher chiral orders and vector-meson contributions were not included. In fact, vector-mesons play an important role in improving the $Q^2$-dependence of the electromagnetic form factors in the large momentum transfer region. In this work, for  completeness, we systematically investigate the pion-loop corrections  to the nucleon electromagnetic form factors up to  $\mathcal{O}(p^4)$ in the  $\mathrm SU(2)$ nonlocal chiral framework and investigate the contributions of vector mesons.

This paper is  organized as follows. In Sec.~\ref{sec2}, we review the basic formulation of the nucleon electromagnetic form factors. In Sec.~\ref{sec3}, we briefly review the construction of the nonlocal chiral Lagrangians that will be used to calculate loop corrections to the nucleon electromagnetic form factors. In Sec.~\ref{sec4}, we explicitly present the loop corrections for all Feynman diagrams up to $\mathcal{O}(p^4)$. In Sec.~\ref{sec5}, we first refit the second, third and fourth-order LECs, as well as the vector meson coupling constants, to the nucleon magnetic moments, electromagnetic radii, and the Dipole and Galster parameterizations of the form factors, then compare the model results for the $Q^2$-dependence of the nucleon electromagnetic form factors with experimental data. Finally, we discuss the contributions of different chiral orders and the fourth moments of the electromagnetic form factors. In the final section, we briefly summarize our results.      
\section{Formulation}
\label{sec2}
The electromagnetic form factors of a spin-$\frac{1}{2}$ particle are defined in terms of the matrix element of electromagnetic vector current as, 
\begin{eqnarray}
  \langle N(p',s') \vert J^{\mu}\vert N(p,s) \rangle  &=&\Gamma^\mu(p,q)=e\bar  u(p+q) [\gamma^\mu  F_1(Q^2)\nonumber \\
  &&+i\frac{\sigma^{\mu\nu}q^\nu }{2m}F_2(Q^2)]u(p),
\end{eqnarray}
where $\Gamma^\mu(p,q)$ denotes the nucleon-photon vertex function, $m$ is the  physical mass of the nucleon, $q$ is the four-momentum transfer with $q^2=-Q^2$, and the Lorentz-invariant  scalar functions $F_1(Q^2)$ and $F_2(Q^2)$ are called the Dirac and Pauli form factors, respectively. For the proton and neutron, $F_1(Q^2)$ and $F_2(Q^2)$ are normalized as $F^p_1(0)=1$, $F^p_2(0)=\kappa_p=1.793$, $F^n_1(0)=0$ and $F^n_2(0)=\kappa_n=-1.913$, where $\kappa$ represents the nucleon anomalous magnetic moments. In the nonrelativistic framework, the matrix element of the electromagnetic vector current is parameterized in terms of the electric and magnetic Sachs form factors $ G_E(Q^2)$ and $ G_M(Q^2)$, which are related to $F_1(Q^2)$ and $F_2(Q^2)$ via,
\begin{eqnarray}
   G_E(Q^2)&=& F_1(Q^2)+\frac{Q^2}{4m^2} F_2(Q^2),\notag\\
  &&\hspace*{-1.8cm} G_M(Q^2)= F_1(Q^2)+ F_2(Q^2).
   \label{eq:vector}
\end{eqnarray}
To characterize the distributions of charge and  magnetic moment, it is necessary to expand the nucleon electromagnetic form factors in  polynomial form \cite{Borah:2020gte,Horbatsch:2016ilr},
\begin{equation}
\begin{split}
G_E^N(Q^2) = G_E^N(0) - \frac{\langle r^2 \rangle_E^N}{6} Q^2 + \frac{\langle r^4 \rangle_E^N}{120} Q^4 + \cdots,\\
G_M^N(Q^2) = G_M^N(0) \left( 1 - \frac{\langle r^2 \rangle_M^N}{6} Q^2 + \frac{\langle r^4 \rangle_M^N}{120} Q^4 + \cdots \right).\\
\end{split}
\end{equation}
where $\langle r^n \rangle$ represents the n-th moments of the electromagnetic form factors. For example, the second moments describe the charge and magnetic radii and are defined in terms of the slope of the form factors at $Q^2=0$,
\begin{eqnarray}
\langle(r_E^p)^2\rangle &=& -\frac{6}{G_E^p(0)}\frac{dG_E^p(Q^2)}{dQ^2},\nonumber \\
\langle(r_M^p)^2\rangle &=& -\frac{6}{G_M^p(0)}\frac{dG_M^p(Q^2)}{dQ^2}, \nonumber\\
\langle(r_E^n)^2\rangle &=& -\frac{6dG_E^n(Q^2)}{dQ^2},\nonumber \\
\langle(r_M^n)^2\rangle &=& -\frac{6}{G_M^n(0)}\frac{dG_M^n(Q^2)}{dQ^2}. \nonumber\\
\label{eq:vector}
\end{eqnarray}
Furthermore, the fourth moments of the electromagnetic form factors define the curvatures of the form factors at $Q^2=0$,   
\begin{eqnarray}
&&\langle(r_E^p)^4\rangle= \frac{120}{G_E^p(0)}\frac{dG_E^p(Q^4)}{dQ^4},\langle(r_M^p)^4\rangle = \frac{120}{G_M^p(0)}\frac{dG_M^p(Q^2)}{dQ^4},\nonumber\\
&&\langle(r_E^n)^4\rangle =\frac{120dG_E^n(Q^2)}{dQ^4},\langle(r_M^n)^4\rangle= \frac{120}{G_M^n(0)}\frac{dG_M^n(Q^2)}{dQ^4}.
\label{eq:vector6}
\end{eqnarray}
In the loop calculation, it can be assumed that there  are two projection operators $\Lambda^1_\mu (p,q)$ and $\Lambda^2_\mu (p,q)$, which project out the form factors $F_1$ and $F_2$ from the vertex operator $\bar \Gamma^\mu(p,q)$ \cite {Knecht:2001qf,Czarnecki:1996rx,Brodsky:1966mv},
\begin{eqnarray}
F_1(q^2) &=& \operatorname{Tr}\left[\Lambda^1_\mu(p,q) \, \bar\Gamma^\mu(p,q)\right], \nonumber\\
F_2(q^2) &=& \operatorname{Tr}\left[\Lambda^2_\mu(p,q) \, \bar\Gamma^\mu(p,q)\right]. \label{eq:vector3}
\end{eqnarray}
where the vertex operator 
$\bar \Gamma^\mu(p,q)$ is related to the $\Gamma^\mu(p,q)$ by $\Gamma^\mu(p,q) =\bar u(p') \bar \Gamma^\mu(p,q) u(p)$. From Eq.~(\ref{eq:vector3}), one can readily derive these  projection operators as,
\begin{widetext}
\begin{eqnarray}
   \Lambda^1_\mu (p,q) &=&  (\centernot p+m)\Big[\frac {1}{4(q^2-4m^2)}{\gamma}_\mu +\frac {3m}{2(q^2-4m^2)^2}(2p_\mu +q_\mu) \Big]  (\centernot p'+m),\notag\\
  &&\hspace*{-1.8cm}\Lambda^2_\mu (p,q) = (\centernot p+m)\Big[-\frac {m^2}{q^2(q^2-4m^2)}{\gamma}_\mu -\frac {m (q^2+2m^2)}{q^2(q^2-4m^2)^2}(2p_\mu +q_\mu) \Big]  (\centernot p'+m).
   \label{eq:operator}
\end{eqnarray}
\end{widetext}
with the on-shell condition $p^2=m^2$. With the help of Eqs.~(\ref{eq:vector3}) and (\ref{eq:operator}), one is able to extract the form factors $F_1(q^2)$ and $F_2(q^2)$ without explicitly analyzing the gamma matrix structure of $\Gamma^\mu(p,q)$. Moreover, many high energy physics program packages have been developed that include some useful tools such as trace calculations and one-loop integrals. For convenience, we will use Mathematica \textbf{\rm{Package-X} } during the calculation \cite{Patel:2015tea}.
\section{nonlocal chiral Lagrangians}

\label{sec3}
\subsection{Nucleon-pion nonlocal chiral Lagrangians}

Since the pion and nucleon are the basic building blocks of  $\rm SU(2)$ chiral perturbation theory, we start by reviewing their chiral transformations. Under the $\rm SU(2)$ chiral transformation, they transform as,
\begin{equation}
     U'(x)= R(x)  U(x)  L^\dag (x), 
     N '(x)=K(x)N (x),
    \label{eq:1}
  \end{equation}
 where $N(x)$ represents the nucleon 
 field and is defined as $N(x)=[n(x),p(x)]^{T}$, and $U(x)$ denotes the $\rm SU(2)$ representation of the pion field and  is given by $U(x)={\rm exp} (i \phi(x)/f)$ with the  pion field matrix $\phi(x)$,
\begin{equation}
\phi(x) =\vec{\tau}.\vec{\pi}(x) = \left( {\begin{array}{*{20}c}
   {\pi ^0(x) } & {\sqrt 2 \pi ^ +(x)  }  \\
   {\sqrt 2 \pi ^ -(x)  } & { - \pi ^0(x) }  \\
\end{array}} \right).
 \end{equation} 
As discussed in Ref.~\cite{Salamu:2025cjv}, the nonlocal chiral Lagrangians for the pion-nucleon interaction can be readily obtained by replacing the pion field operator $U(x)$ with an extended nonlocal pion field $\int d^4a F_{\pi}(a) U(x+a)$, where $F_{\pi}(a)$ is a scalar regulator function for the  pion. Nevertheless, the chiral transformation of the nonlocal pion fields does not obey the chiral transformation rules of the local pion field in Eq.~(\ref{eq:1}). To recover the chiral transformation of the nonlocal fields, it needs to be modified by adding a gauge link operator. Thus, the total nonlocal operators can be defined as~\cite{Salamu:2025cjv},
\begin{eqnarray}
\hat  U(x)&=& {\rm exp} {\bigl \{} \frac{1}{c_0}\int d^4a F_{\pi}(a) {\rm Log} {\bigl [}G_R(x,x+a)U(x+a) \nonumber \\
&& G^\dag_L(x,x+a) {\bigr]} {\bigr \}},
\label{eq:10}
\end{eqnarray}
where $c_0$ is a normalization constant and satisfies $c_0\equiv\int da F_{\pi}(a)$, $G_R(x,{x+a})$ and $G_L(x,{x+a})$ represent the Wilson line operators for the left and right-handed gauge fields and are conventionally parameterized as, 
\begin{eqnarray}
  G_R(x,{x+a})&=&{\cal{P}} {\rm exp}\left[-i  \int_{x}^{x+a} dz^\mu  r_\mu (z) \right],\nonumber\\
  G_L(x,{x+a})&=&{\cal{P}}{\rm exp}\left[-i \int_{x}^{x+a} dz^\mu  l_\mu (z)  \right],
\label{eq:9}
\end{eqnarray}
where ${\cal{P}}$ denotes the path-ordering operator. It can be  proved that under the chiral transformation, the left and right handed Wilson line operators transform as,
\begin{equation}
\begin{split}
G'_{L[R]}(x,y)= L(x)[ R(x)]G_{ L[R]}(x,y){ L}^\dag(y)[ R^\dagger(y)].
\end{split}
\label{eq:44}
\end{equation}
Using these Wilson line operators, one can derives that the nonlocal pion field operator $\hat U(x)$ transforms approximately as in Eq.~(\ref{eq:1})~\cite{Salamu:2025cjv}. In particular, under the $\rm SU_V(2)$ transformation [$L(x)=R(x)=V(x)$], the nonlocal pion field exactly transforms as $\hat U'(x)=V(x)\hat U(x)V^\dagger(x)$. This implies that the nonlocal Lagrangian for the pion-nucleon interactions can be constructed by replacing the original local pion field $U(x)$ with the nonlocal one $\hat U(x)$. For example, the lowest order Lagrangian for pion-nucleon interactions is given by \cite{Fettes:1998ud}, 
\begin{equation}
    {\cal L}_{\pi N}^{(1)} = \bar{\Psi} \Bigl(
i \tilde{ \centernot D} - m_0 + \frac{g_A}{2} u_\mu \gamma^\mu \gamma_5 \Bigr) \Psi,
\label{eq:e1}
\end{equation}
where $m_0$ is the nucleon mass in the chiral limit, $g_A$ is the nucleon axial charge in the chiral limit, the nucleon covariant derivative $D_\mu$ is defined as
$D_\mu = \partial_\mu + \Gamma_\mu$ with the nonlocal vector and axial vector connection $\Gamma_\mu$ and  $ u_\mu$, 
\begin{eqnarray}
\Gamma_\mu
&=& \frac{1}{2}\left( \hat{u}^\dagger \partial_\mu \tilde{u} + \hat{u} \partial_\mu \hat{u}^\dagger
    \right)-\frac{i}{2} \left( \hat{u}^\dagger Q \hat{u} + \tilde{u} Q \hat{u}^\dagger
    \right) A_\mu,			\nonumber		\\
u_\mu
&=& 
    i\left( \hat{u}^\dagger \partial_\mu \hat{ u} - \hat{u} \partial_\mu \hat{u}^\dagger
    \right)
 +
    \left( \hat{u}^\dagger Q \hat{u} - \hat{u} Q \hat{u}^\dagger
    \right) A_\mu,
\label{eq:22}
\end{eqnarray}
where the nonlocal $\hat u$ is defined as $\hat u=\sqrt{\hat{\rm U}}$, $\hat u^\dag=\sqrt{\hat{\rm U}^\dag}$, $A_\mu$ represents the photon field, and $Q$ denotes the nucleon charge matrix and takes the form  $Q=-e\, \rm diag(1,0)$.  
At the second order, the pion-nucleon interactions  include terms proportional to LECs $c_i$ $(i=1-7)$, where only $c_2$, $c_4$, $c_6$ and $c_7$ related terms contribute to the electromagnetic form factors and are given by\cite{Fettes:1998ud}, 
\begin{widetext}
\begin{equation}
  {\cal L}_{\pi N}^{(2)} = \bar{\Psi} \, \biggl\{
-\frac{c_2}{8m^2} \Bigl( \langle u_\mu u_\nu \rangle \{ D^\mu,D^\nu \} +{\rm h.c.} \Bigr)
+\frac{i\,c_4}{4} \sigma^{\mu\nu} [ u_\mu,u_\nu] +\frac{c_6}{8m} \sigma^{\mu\nu} F^+_{\mu\nu}
+\frac{c_7}{8m} \sigma^{\mu\nu} \langle F^+_{\mu\nu} \rangle
\biggr\} \, \Psi, \\
\label{eq:e2}
\end{equation}
\end{widetext}
where the chiral covariant field strength $F_{\mu\nu}^+ = {\hat u}^\dagger F_{\mu\nu} {\hat u} + {\hat u} F_{\mu\nu} {\hat u}^\dagger$, with the photon field strength tensor $F_{\mu\nu} =Q (\partial_\mu A_\nu - \partial_\nu A_\mu)$. As for the third and fourth order chiral Lagrangian, we only list the terms that contribute to the nucleon electromagnetic form factors\cite{Fettes:1998ud},
\begin{widetext}
\begin{eqnarray}
{\cal L}_{\pi N}^{(3)} &=& \bar{\Psi} \, \biggl\{
\frac{i\,d_6}{2m}\Bigl( [D^\mu,\hat{F}^+_{\mu\nu}]D_\nu +{\rm h.c.} \Bigr)
+\frac{i\,d_7}{2m}\Bigl( [D^\mu,\langle F^+_{\mu\nu}\rangle]D_\nu +{\rm h.c.} \Bigr)
\biggr\} \, \Psi, \nonumber \\
{\cal L}_{\pi N}^{(4)} &=& \bar{\Psi} \, \biggl\{
-\frac{e_{54}}{2} \bigl[ D^\lambda,[D_\lambda, \langle F^+_{\mu\nu} \rangle ]\,]\,\sigma^{\mu\nu}
-\frac{e_{74}}{2} \bigl[ D^\lambda,[D_\lambda, \hat{F}^+_{\mu\nu} ]\,]\,\sigma^{\mu\nu}-\frac{e_{105}}{2}\langle F^+_{\mu\nu} \rangle \, \langle \chi_+ \rangle \, \sigma^{\mu\nu}-\frac{e_{106}}{2}\hat{F}^+_{\mu\nu} \langle \chi_+ \rangle \, \sigma^{\mu\nu}
\biggr\} \,\Psi,\nonumber \\
\label{eq:e3}
\end{eqnarray}
\end{widetext}
where the traceless operator $\hat{A}$ is defined as $\hat{A} = A -\frac{1}{2}\langle A \rangle$, and the chiral covariant mass term $\chi_\pm$ is given by $\chi_\pm = \hat u^\dagger \chi \hat u^\dagger \pm \hat u \chi^\dagger \hat u$, where the mass field  $\chi $ is expressed in terms of linear combination of the scalar and pseudoscalar fields via $\chi = 2B(s + i p)$. The constants $d_6$, $d_7$, $e_{54}$ and $e_{74}$ are the third and fourth order LECs, which usually can be fitted to the electromagnetic charge radii of the proton and neutron.  
\subsection{Nucleon-vector meson interaction}

It is well established that, the vector meson pole contribution plays an important role in describing the $Q^2$-dependence of the nucleon form factors, especially when one investigates the distributions of the nucleon electromagnetic form factors in the large $Q^2$-region. To this end, we need to include the vector mesons in the nonlocal chiral perturbation theory. To construct the nonlocal interaction between the vector meson and nucleon, by analogy with the pion-nucleon interaction, the vector meson field operator $V_\mu(x)$ can be  replaced  with the nonlocal one  $\int  d^4a F_V(a) V_\mu(x+a)$. Nevertheless, the nonlocal vector field operator also breaks the rule of local chiral transformation. We omit the explicit details of the chiral transformation of the nonlocal vector meson and  the corresponding gauge link operator, since they do not introduce additional contributions to the nucleon electromagnetic form factors. Following Refs.~\cite{Machleidt:2000ge,Nozawa:1989pu,Faessler:2007bc}, we construct the leading-order nonlocal vector meson-nucleon interaction as, 
\begin{widetext}
\begin{eqnarray}
{\cal L}^{(1)}_{ NV}&=&\int F_{V}(a) d^4a [{\rm g}_{e\rho}  \bar N {\gamma}^{\mu}  {\bm \tau}.{\bm \rho}_\mu(x+a)   N-\frac{{\rm g_{m\rho}}}{2 m}  \bar N  \sigma^{\mu\nu} \partial_\nu {\bm \tau}.{\bm \rho}_\mu(x+a) N+{\rm g_{e\omega}}  \bar N {\gamma}^{\mu}   \omega_\mu(x+a)   N\nonumber \\
&&-\frac{{\rm g_{m\omega}}}{2 m}  \bar N  \sigma^{\mu\nu} \partial_\nu \omega_\mu(x+a) N],
\label{eq:18}
\end{eqnarray}
\end{widetext}
where $\rm g_{e\rho}$, $\rm g_{e\omega}$, $\rm g_{m\rho}$ and $\rm g_{m\omega}$ represent the vector and tensor coupling constants between the nucleon and vector mesons, and the $\rho$ vector-meson  matrix is parameterized as, 
\begin{equation}
{\bm \tau}.{\bm \rho}_\mu = \left( {\begin{array}{*{20}c}
   {\rho_\mu ^0 } & {\sqrt 2 \rho ^ +  }  \\
   {\sqrt 2 \rho_\mu ^ -  } & { - \rho_\mu ^0 }  \\
\end{array}} \right).
 \end{equation}
 Note that the nonlocal interaction in Eq.~(\ref{eq:18}) naturally generates an extra regulator $\widetilde{F}_V (k)$, whose role has been investigated in early works on nucleon-nucleon scattering  \cite{Chiang:2001as,Machleidt:2000ge,Machleidt:1995km,Machleidt:1987hj}. In addition to this, the Lagrangian for leading-order vector meson-photon interaction is given by \cite{HillerBlin:2017syu},
\begin{equation}
{\cal L}^{(\rm 2)}_{V \gamma}=-\frac{eF_\rho}{2m_\rho}   \rho_{\mu\nu}^{0}  f^{\mu\nu}-\frac{eF_\omega}{2m_\omega}  f^{\mu\nu} \omega_{\mu\nu},
\label{eq:20}
\end{equation}
where $m_\rho$ and $m_\omega$ are the masses of the vector mesons, $F_\rho$ and $F_\omega$ represent the vector meson decay constants and can be determined from the dilepton decay widths via  $\Gamma(V\rightarrow ee^- )=\frac{4\pi \alpha^2 F^2_V }{3 m_V}$ \cite{HillerBlin:2017syu,HillerBlin:2018gjw,Klingl:1996by, Maris:1999nt}. Using the vector meson masses $m_\rho=770\ \rm MeV$, $m_\omega=782\ \rm MeV$ and the latest data for the dilepton decay widths $\Gamma(\rho\rightarrow ee^- )=6.96\ \rm keV$ and $\Gamma(\omega\rightarrow ee^- )=0.64\ \rm keV$\cite{ParticleDataGroup:2024cfk}, we obtain the vector meson decay constants as $F_\rho=155\ \rm MeV$, $F_\omega=47.37 \ \rm MeV$, respectively. By contracting the vector mesons in Eqs.~(\ref{eq:18}) and~(\ref{eq:20}), one can obtain the vector meson pole diagram, which contains a vector meson propagator. In the EFT, the dynamics of the vector meson can be described in  two distinct ways: the vector and the tensor representations. This, in turn, gives rise to two different forms of the vector meson propagator. In this work, we make use of the vector field representation of the vector meson, and the propagator for the massive vector field is given by \cite{Borasoy:1995ds},
\begin{equation}
G^{\mu\nu}(q)=\frac{i (-g^{\mu \nu}+\frac{q^\mu q^\nu }{m_V^2})}{q^2-m_V^2 +i\varepsilon}.
\end{equation}

\section{Nucleon electromagnetic from factors  }
\label{sec4}
\begin{figure*}
\centering
\includegraphics[width=0.9\textwidth]{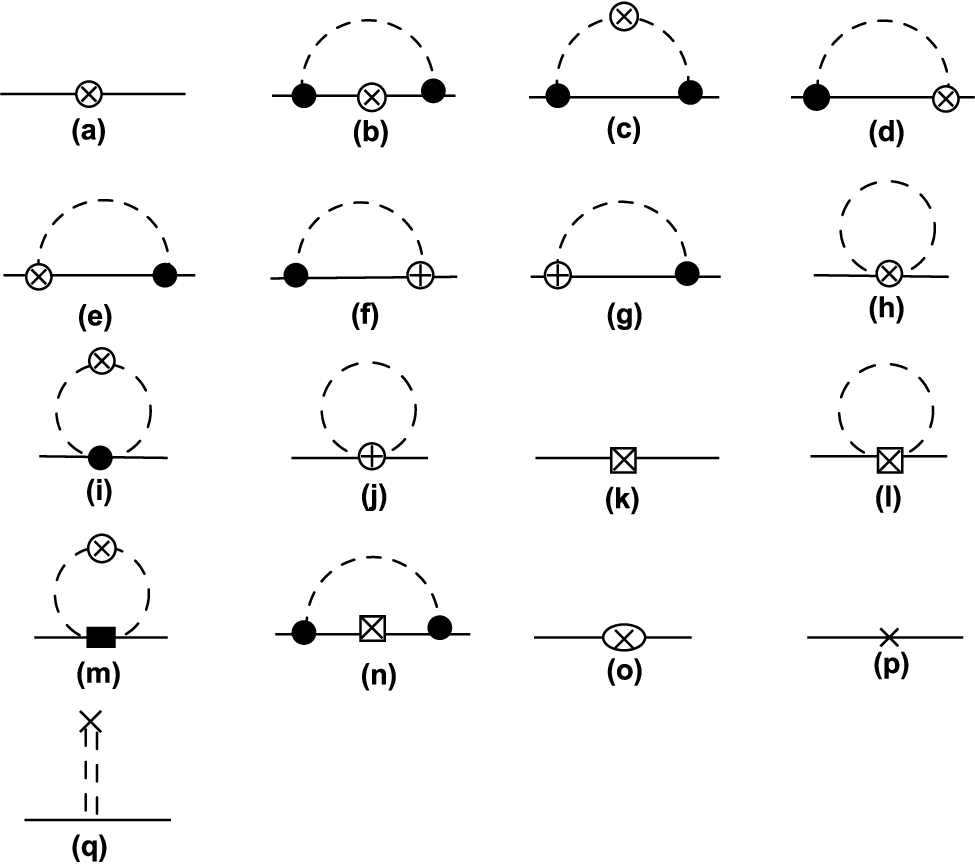}
\caption{The Feynman diagrams arise from the first, second, third and  fourth  order chiral interactions of Eqs.~(\ref{eq:e1}), (\ref{eq:e2}) and (\ref{eq:e3}), where the solid, dashed and double-dashed lines represent the pion, nucleon and vector-meson fields, the solid dots and squares denote the first and second order pion-nucleon strong coupling vertices, the cross with circle and square are the lowest and second order photon-baryon interactions, the circled  plus refers to the  additional gauge link vertex, the cross and cross with ellipse are the third and fourth order photon-nucleon interactions.}
\label{fig:loop}
\end{figure*}

In this section, we explicitly present the pion one-loop corrections to the electromagnetic form factor of the nucleon. The relevant Feynman diagrams are displayed in Fig.~(\ref{fig:loop}), where Figs.~\ref{fig:loop}(a) to \ref{fig:loop}(j) represent the lowest chiral order  contributions arising from the Lagrangian in Eqs.~(\ref{eq:e1}), Figs.~\ref{fig:loop}(k) to \ref{fig:loop}(n) are the second chiral order contributions from Eq.~(\ref{eq:e2}), Figs.\ref{fig:loop}(o), and \ref{fig:loop}(p) represent the third and fourth order contributions arising from the Lagrangian in Eq.~(\ref{eq:e1}), while  Fig.~\ref{fig:loop}(s) represents the vector meson contribution. The explicit contributions of each Feynman diagram to the electromagnetic vertex function $\Gamma(p,q)$  are listed in App. \ref{sec.6}. 

Note that at lowest order, the tree level-diagram contribution is identical to that in the local case and only contributes to the electric form factor $G_E(Q^2)$ of the proton. As for the loop  diagrams, they contain a nonlocal regulator function $\widetilde{F}_{\pi}(k)$ due to the pion nonlocality. Consequently, the loop contributions are convergent by virtue of $\widetilde{F}_{\pi}(k)$, which modifies the superficial degree of divergence of the loop integral. Moreover, as  mentioned in the previous section, the gauge link operators of Eq.~(\ref{eq:9}) give rise to additional loop contributions as shown in  [Fig.\ref{fig:loop}(f+g)] and [Fig.\ref{fig:loop}(j)], which guarantee the gauge invariance of the electromagnetic interaction. Analytically, these contributions include the derivative of the nonlocal regulator $\widetilde{F}_{\pi}(k)$, which vanishes in the local limit $\widetilde{F}_{\pi}(k)\rightarrow 1$. As a result, the loop corrections reduce to  those of the local case. 

At the second order, the tree diagram contributes to the magnetic form factors of the neutron and proton, and the resulting the vertex function is identical to that of the  local case. In addition to this, at the second chiral order, we also include pion loop corrections as displayed in Figs.~\ref{fig:loop}(l)-(n). Compared with the first-order contributions, the second-order nonlocal Lagrangian does not produce additional gauge-link diagrams, but the loop integrals are UV convergent as desired.     

As for the third and fourth orders contributions, we only take into account the tree diagrams, which contribute to the both $F_1(Q^2)$ and $F_2(Q^2)$ form factors and are proportional to the four-momentum transfer $Q^2$. Therefore, the electromagnetic radii of the nucleon, which are defined in terms of the slopes of the electromagnetic form factors at $Q^2=0$, receive non-negligible contributions from these chiral orders. Finally, as displayed in Fig.\ref{fig:loop}(s), the photon couples to the nucleon via a vector-meson propagator. The corresponding the vector-meson vertex function, as shown in Eq.~(\ref{eq:A15}), contains an additional regulator $\widetilde{F}_{V}(k)$. As a result, the local vector meson pole is rapidly suppressed and generates adequate curvatures for the electromagnetic form factors. 

\section{Numerical results}

\label{sec5}
\subsection{Fitting of LECs and cut-of mass}

As demonstrated in the previous section, within the  nonlocal framework the loop contributions to the electromagnetic form factors contain a nonlocal regulator. To calculate the loop integrals, it is necessary to define the explicit expression of the regulator function of the pion and vector meson $\widetilde{F}_\phi (k)$($\phi=\pi,V$). In general, $\widetilde{F}_\phi (k)$ is an analytic function of $k^2$ and can usually  be parameterized in terms of Gaussian and dipole form. Following Refs.~\cite{Forkel:1994yx, Musolf:1993fu}, we choose 
the regulator to have a simple dipole shape in $k^2$,
\begin{equation}
\widetilde{F}_\phi (k)
= \left( \frac{m_\phi^2 -\Lambda_\phi^2}{k^2-\Lambda_\phi^2+i\varepsilon } \right)^2,
\label{eq:re}
\end{equation}
where  $m_\phi^2 $ denotes the masses of the pion and vector-mesones, and $\Lambda_\phi$ is the cut-off mass of the  dipole regulator. 
For the pion cut-off mass, following  Refs.~\cite{Salamu1,Salamu2}, $\Lambda_\pi$ was determined as $\Lambda_\pi=(1.0\pm0.1)$GeV. As for the vector meson, we assume that the vector mesons have a common cutoff mass    $\Lambda_\rho=\Lambda_\omega$.

Apart from this, at the second chiral order, the loop corrections involve  second order LECs $c_1$, $c_2$, $c_3$ and $c_4$. However, since they  depend on the renormalization scheme, their values  in the local framework cannot be directly applied here. As shown in Ref.~\cite{Salamu:2025cjv}, in nonlocal framework, they can be fitted to the pion mass dependence of the nucleon mass $m_N$ and axial charge $g_A$ and take  the values of $c_1=(-0.211\pm0.020){\rm GeV^{-1}}$, $c_2 =(-0.221 + 0.021){\rm GeV^{-1}}$, $ c_3=(-0.055\pm 0.084){\rm GeV^{-1}}$, $c_4=(1.095\pm0.128){\rm GeV^{-1}}$. As for the  remaining second order LECs $c_6$ and $c_7$, we follow the standard  procedure and fit them to the proton and neutron magnetic moments $\mu_p=2.793$ and $\mu_n=-1.913$. The fit yields $c_6=6.263$ and $c_7=-2.979$, respectively. By comparison, we find that the absolute values of second-order chiral coupling constants $c_6$  and $c_7$  are considerably larger  than their local counterparts \cite{Kubis:2000zd}.

At the third and fourth chiral orders, we only take into account the tree diagrams, which are  proportional to LCEs $d_6$, $d_7$, $e_{74}$ and $e_{54}$. Indeed, without the vector meson contribution, these parameters can be accurately determined from  the electromagnetic radii of the proton and neutron. However, the inclusion of the vector meson introduces additional free parameters--nucleon-vector meson strong coupling constants $\rm g_{e\rho}$, $\rm g_{e\omega}$, $\rm g_{m\rho}$, $\rm g_{m\omega}$ and the vector meson dipole cut-off mass $\Lambda_\rho$. To constrain the number of parameters, we adopt the latest experimental values of the nucleon electromagnetic radii $\langle r^2\rangle_E^p=0.8409^2[\rm fm^2]$, $\langle r ^2\rangle_M^p=0.851^2[\rm fm^2]$, $\langle r ^2\rangle_E^n=-0.1155 \ [\rm fm^2]$ and $\langle r^2\rangle_M^n=0.864^2 \ [\rm fm^2]$ \cite{ParticleDataGroup:2024cfk}. Nevertheless, the vector meson coupling constants and the cut-off mass remain unfixed. In fact, they can be fixed from the nucleon-nucleon scattering  and dispersion analysis. Unfortunately, the vector meson coupling constants are poorly constrained across the different fitting procedures and exhibit strong model dependence\cite{HillerBlin:2017syu}. For example, the meson-exchange model of the nucleon-nucleon scattering yields  $1.8\leq \rm g_{e\rho } \leq3.2$, $ 4.3\leq \rm g_{m\rho }/g_{e\rho } \leq 6.6$, $8\leq \rm g_{e\omega }\leq 20 $ and $-1\leq g_{m\omega }/g_{e\omega }\leq 0$, while dispersion analysis yields  $\rm g_{e\rho }=2$, $\rm g_{m\rho }/g_{e\rho } = 6.1$, $\rm g_{e\omega }=20.9$ and $\rm g_{m\omega }/g_{e\omega }=-0.16$ even though the nonlocal effects were taken into account by including the nonlocal vertex regulator function  $\widetilde{F}_\phi (k)$ with the cut-off masses $\Lambda_\rho=1.3$ and $2 \ \rm GeV$ \cite{Chiang:2001as,Mergell:1995bf,Drechsel:1998hk,Machleidt:2000ge}. Although we attempt to extract the vector meson coupling constants from the curvatures (fourth moments) of the electromagnetic form factors, the values of curvature are not accurately determined experimentally (for more details, see Subsec~.C).   

Therefore, in a straightforward manner, we fit them to the $Q^2$-dependence of the nucleon electromagnetic form factors, such as dipole and Galster-type forms, rather than directly to the  experimental data, since the parameterizations provide a smooth and analytic representation of the data and  reduce the sensitivity to individual data fluctuations. Generally, the nucleon electromagnetic form factors can be parameterized in dipole and Galster-type forms as \cite{Atac:2021wqj,Alexandrou:2017ypw},  
\begin{eqnarray}
G^{j}_{i}(Q^2)&=& \frac{G^{j}_{i}(0)}{(1+Q^2/M^2_{ij})^2},\nonumber\\ 
G^{E}_{n}(Q^2)&=&\frac{1}{(1+Q^2/A)^2}\frac{B \tau}{1+C \tau},
\label{eq:re1}
\end{eqnarray}
where $i=E,M$ and $j=p,n$, $M^2_{ij}$ denotes the cut-off masses of nucleon electromagnetic form factors, $\tau=\frac{Q^2}{4m^2}$ is the  dimensionless variable, with the fitted parameters $A=
(0.505\pm0.079)\rm GeV^2$, $B=1.655\pm0.126$, and $C=0.909\pm0.583$. In the traditional dipole parameterization, a universal cut-off mass $M^2_{ij}=0.71\ \rm GeV^2$ is used. However, such parametrization cannot produce  accurate nucleon electromagnetic radii. To improve the description, we refit them to the latest experimental data for the nucleon electromagnetic charge radii. The fitted values of cut-off mass $M^2_{ij}$ are listed in Tab.~\ref{tab1}. As expected, these values are different from the conventional cut-off mass $M^2_{ij}=0.71 \ \rm GeV^2$. With these updated cut-off masses, and using the dipole and Galster-type parameterization in Eq.~(\ref{eq:re1}) as a data generator, we fit the vector meson-nucleon coupling constants $g_{e\rho}$, $g_{e\omega}$, $g_{m\rho}$,  $g_{m\omega}$ and the vector meson cut-off mass $\Lambda_\rho$ to  $G(Q^2)$ by performing $\chi^2$ analysis.
\begin{table*}
\caption{The cut-off masses  of the nucleon electromagnetic form factors are fitted to the experimental data of the nucleon electromagnetic  radii.}
\begin{center}
\begin{tabular}{c|cccc|ccc|ccc|ccccccc}
  \hline \hline
    &&& $G^p_E $&&& $G^p_M $ &&& $G^n_E $  &&& $G^n_M $ && \\  \hline 
   EM  radii $\langle r^2\rangle^j_i [\rm fm^2]$\cite{ParticleDataGroup:2024cfk} &&& $0.8409^2$&&& $0.851^2$&&&$-0.110$ &&&$0.864^2$  \\ 
\hline
 cut-off mass$M^2_{ij}$ $[\rm GeV^2]$   &&& $0.813^2$ &&& $0.803^2$&&&- &&&$0.791^2$ \\ 
 \hline\hline
\end{tabular}
\end{center}
\label{tab1}
\end{table*}

The least-squares fit analysis yields  $\rm g_{e\rho}=5.241$, $\rm g_{e\omega}=25.197 $, $\rm g_{m\rho }/g_{e\rho }=6.183$, $\rm g_{m\omega }/g_{e\omega }=0.055$, and $\Lambda_\rho=1.79\rm GeV$, respectively. Notably, the fitted values for the vector-to- tensor coupling constant  ratios $\rm g_{m\rho}/g_{e\rho}$ and $\rm g_{m\omega}/g_{e\omega}$ as well as $\Lambda_\rho$ are almost consistent with the ranges extracted from nucleon-nucleon scattering. In contrast, the vector coupling constants $ \rm g_{e\rho}$ and $\rm g_{e\omega }$ are slightly larger than the upper limits of the nucleon-nucleon scattering data ranges. This discrepancy indicates that the nucleon electromagnetic form factors may be  affected by the  massive vector mesons and complicated off-shell background effects. Correspondingly, $\mathcal{O}(p^3)$ and $\mathcal{O}(p^4)$ chiral low energy coupling constants are obtained as $d_6=-0.247{\rm GeV^{-2}}$, $d_7=-0.092{\rm GeV^{-2}}$, $e_{54}=0.041 {\rm GeV^{-3}}$ and $e_{74}=0.438 \ {\rm GeV^{-3}}$. Clearly, these values totally differ in magnitude and sign from those obtained with the IR and EOMS renormalization methods \cite{Kubis:2000zd,Schindler:2005ke}. This reflects the fact that chiral low-energy coupling constants are renormalization scheme-dependent. 

\subsection{The $Q^2$-dependence of nucleon electromagnetic form factors }
\begin{figure*}
\centering
\begin{tabular}{ccc}
\hspace{0.1cm}{\epsfxsize=3.2in\epsfbox{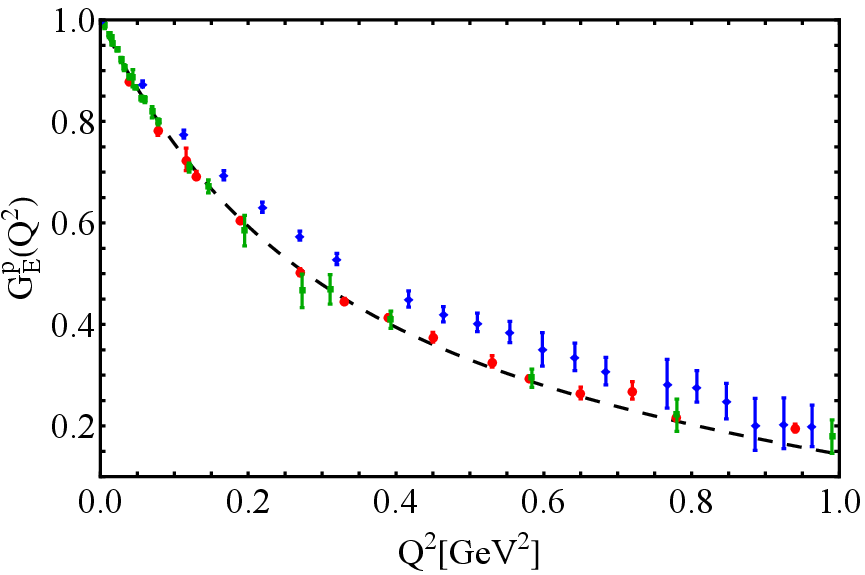}}&
\hspace{0.1cm}{\epsfxsize=3.2in\epsfbox{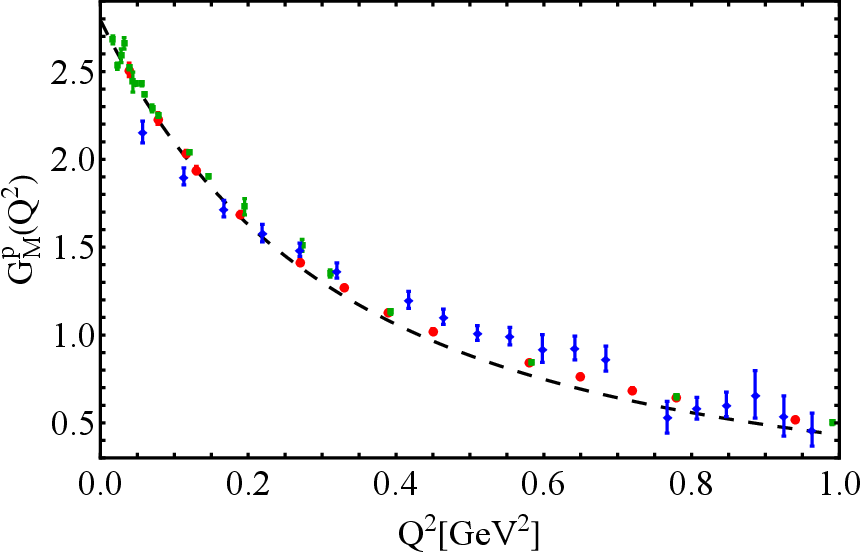}}& \\
\hspace{0.1cm}{\epsfxsize=3.2in\epsfbox{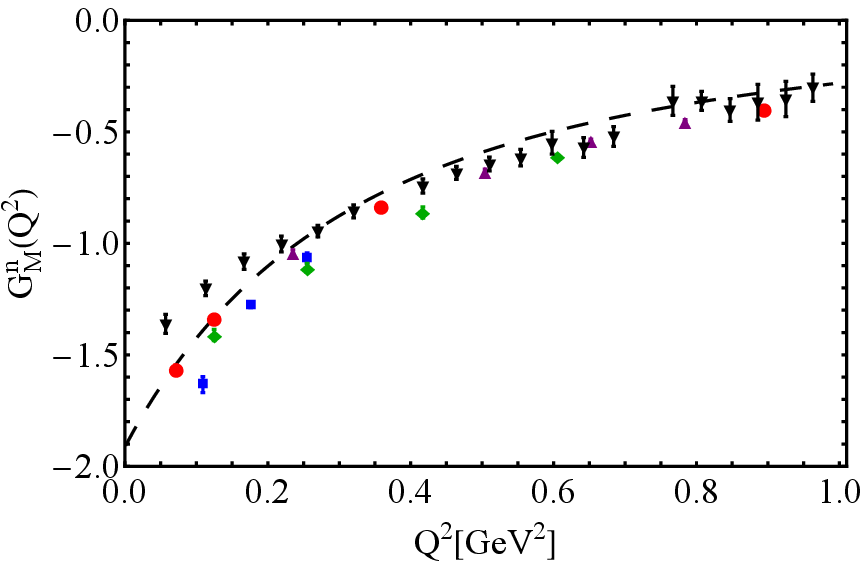}}&
\hspace{0.1cm}{\epsfxsize=3.2in\epsfbox{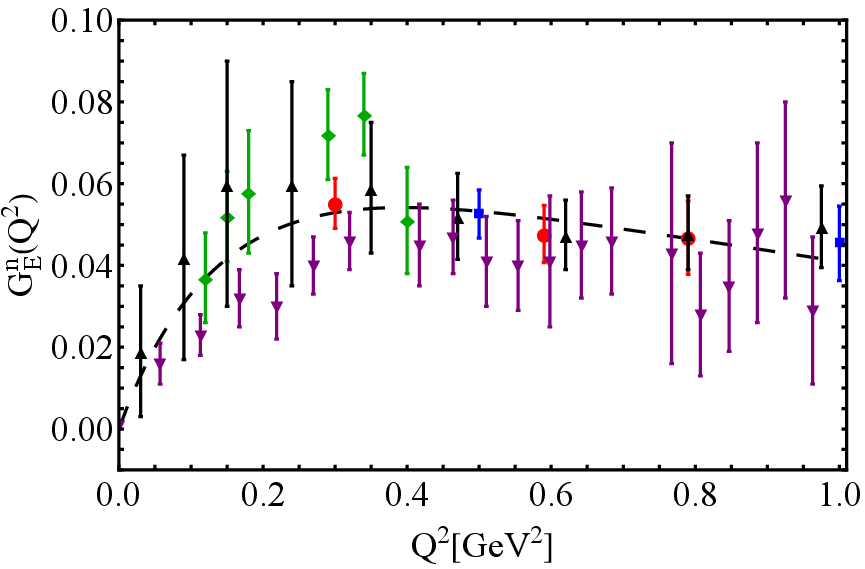}}&
\end{tabular}
\caption{The model results (short-dashed lines) for the proton (left panel) and neutron (right panel)    electromagnetic form factors are compared with the experimental and lattice data from \cite{Price:1971zk,Hohler:1976ax,Glazier:2004ny,JeffersonLabE93-026:2003tty,Herberg:1999ud,Schiavilla:2001qe,Alexandrou:2018sjm}.}
\label{fig:2}
\end{figure*}

\begin{figure*}
\centering
\begin{tabular}{ccc}
\hspace{0.1cm}{\epsfxsize=3.2in\epsfbox{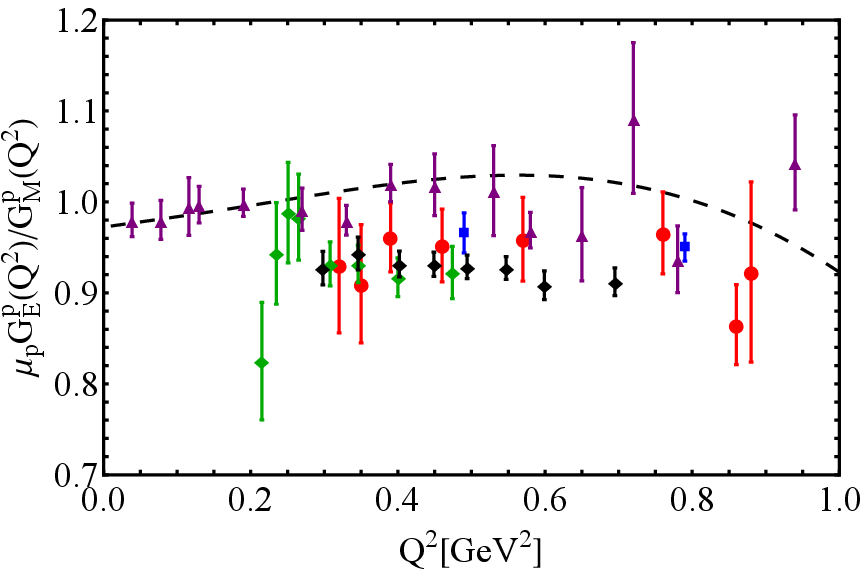}}&
\hspace{0.1cm}{\epsfxsize=3.2in\epsfbox{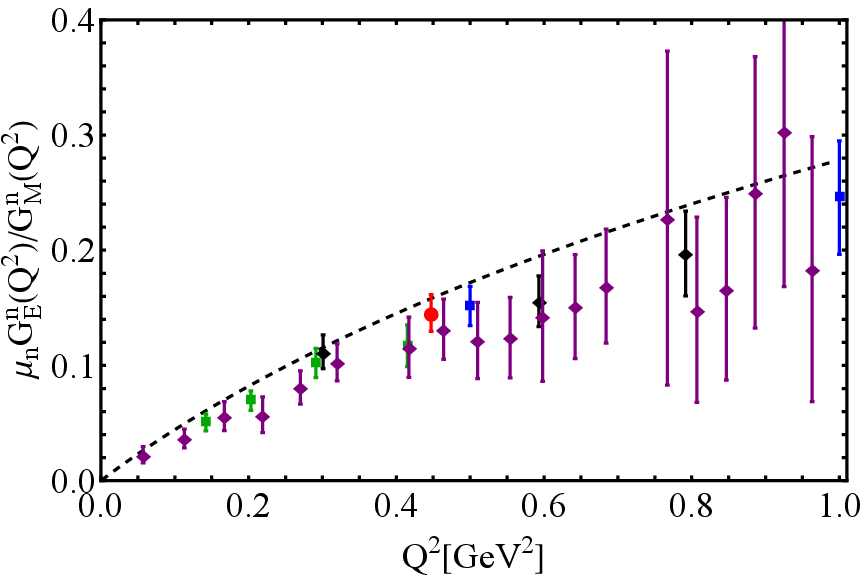}}& 
\end{tabular}
\caption{The model results (short-dashed lines) for the ratio of the proton (left panel) and neutron (right panel) electromagnetic form factors are compared with experimental and lattice  data from \cite{Price:1971zk,Zhan:2011ji,JeffersonLabHallA:2011yyi,JeffersonLabHallA:1999epl,Gayou:2001qt,Alexandrou:2018sjm,Schlimme:2021cdm,BLAST:2008bub,JeffersonLabE93-026:2003tty,JeffersonLaboratoryE93-038:2005ryd}.}
\label{fig:3}
\end{figure*}

With these best-fit coupling constants and cutoff masses, we now proceed to compute the form factors. The nonlocal model predictions for the $Q^2$-dependence of the electromagnetic form factors, in the $0\leq Q^2\leq 1\rm GeV^2$ region, are displayed in Fig.~\ref{fig:2}. From Fig.~\ref{fig:2}, it can be seen that the $Q^2$-dependence of the nucleon electromagnetic form factors at large $Q^2$ region is significantly improved and is consistent with the experimental and lattice data. This is an important feature of the nonlocal formalism, since the higher-order contributions, which govern the high-energy(short-distance) behavior of chiral loops in the local perturbation theory, are  suppressed due to the nonlocal regulator. Actually, a similar conclusions have been reached in earlier analyses of pion and nucleon electromagnetic form factors within the nonlocal quark model \cite{Ivanov:1996pz,Gutsche:2012ze,Faessler:2003yf,Forkel:1994yx, Musolf:1993fu}.     
Similarly, as shown in Fig.~\ref{fig:3}, the model predictions for the ratios of the nucleon electromagnetic form factors are compared with  experimental and lattice  data. The nonlocal results demonstrate that the proton and neutron form factor ratios are consistent with the data over a wide  region $0\leq Q^2\leq 1\ \rm GeV^2$.   
\subsection{Curvatures of nucleon electromagnetic   form factors}
\label{subsec.c}
\label{subsec.1}
As shown in Eq.~(\ref{eq:vector6}), the fourth moments of the nucleon electromagnetic form factors $\langle r^4\rangle$ describe the curvature of $G(Q^2)$ at $Q^2=0$. Mathematically, any perturbative variation of the $\langle r^4\rangle$ will cause a drastic change of the $Q^2$-distributions  of the  electromagnetic form factors. More importantly, this also leads to a wide range of uncertainty in the  electromagnetic radii since the electron-nucleon scattering cannot reach $Q^2=0$ \cite{Horbatsch:2016ilr,Yan:2018bez, Sick:2017aor}. To handle this issue, it is necessary to accurately fix the $\langle r^4\rangle$ as well as other higher order moments of form factors. 

However, as listed in Tab.~\ref{tab2}, the curvatures of nucleon electromagnetic form factors explicitly depend on the functional form of $G(Q^2)$ and have a wide uncertainty range across the different parameterizations. For example, the values $\langle r^4\rangle$ from the standard dipole-Galster parameterization in Eq.~(\ref{eq:re1}) are smaller than those from the  polynomial and  rational function parameterization  \cite{Alarcon:2018irp,Kelly:2004hm,Alberico:2008sz,Borah:2020gte}. In fact, the dipole type form of $G(Q^2)$ decreases more quickly than the experimental data  \cite{Cloet:2010qa}. Thus, the curvatures obtained from both dipole and nonlocal methods are close to each other since the higher energy contributors of meson loops is suppressed due to the nonlocal regulator. This further highlights the model dependence of the extraction procedure.      
\begin{table*}
\caption{The curvatures of the proton and neutron  electromagnetic form factors from different parameterization groups.}
\begin{center}
\begin{tabular}{c|cccc|ccc|ccc|ccccccc}
  \hline \hline
  fit  &&& $\langle r^4\rangle^p_E[\rm fm^4] $&&& $\langle r^4\rangle^p_M [\rm fm^4]$ &&& $\langle r^4\rangle^n_E[\rm fm^4] $  &&& $\langle r^4\rangle^n_M[\rm fm^4] $ && \\  \hline 
  Ref. \cite{Borah:2020gte} &&& $1.08\pm 0.284$ &&& $-2.0\pm1.88$&&&$-0.33\pm0.242$ &&&$-2.3\pm 2.37$ \\ 
  \hline
    Ref.\cite{Kelly:2004hm} &&& $1.621\pm0.064$ &&& $1.471\pm0.034$&&&$-0.330\pm0.016$ &&&$2.324\pm0.748$ \\ 
  \hline
  Ref. \cite{Alberico:2008sz} &&& $1.620\pm0.050$ &&& $1.720\pm0.025$&&&$-0.330\pm0.140$ &&&$2.920\pm2.600$ \\ 
  \hline
   Ref.\cite{Alarcon:2018irp} &&& $1.536\pm0.065$ &&& $1.729\pm0.053$&&&$-0.571\pm0.065$ &&&$2.044\pm0.002$ \\ 
  \hline
   Dipole\&Galster  &&& $1.250$ &&& $1.311$&&&$-0.361$ &&&$1.393$ \\ 
\hline
 This work  &&& $1.331$ &&& $1.380$&&&$-0.422$ &&&$1.516$ \\ 
 \hline\hline
\end{tabular}
\end{center}
\label{tab2}
\end{table*}
\subsection{Contributions from different chiral orders}
\begin{figure*}
\centering
\begin{tabular}{ccc}
\hspace{0.1cm}{\epsfxsize=3.2in\epsfbox{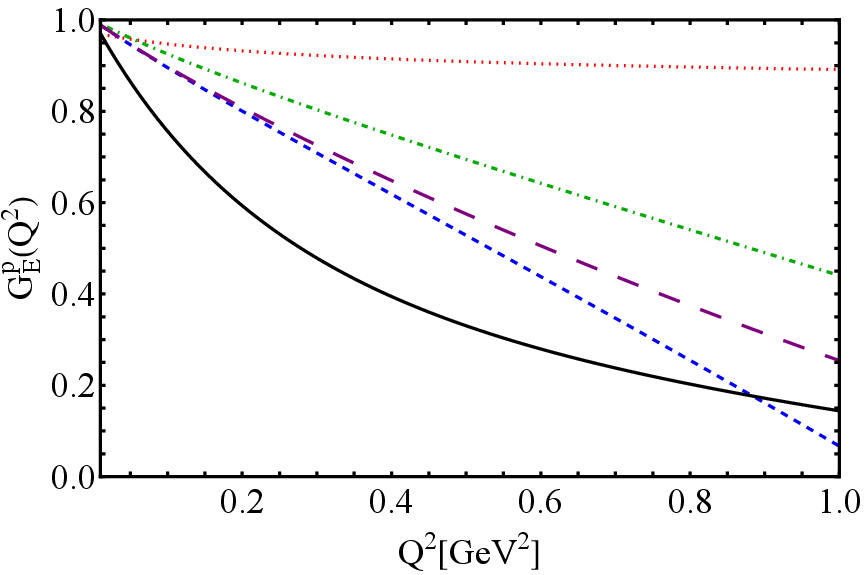}}&
\hspace{0.1cm}{\epsfxsize=3.2in\epsfbox{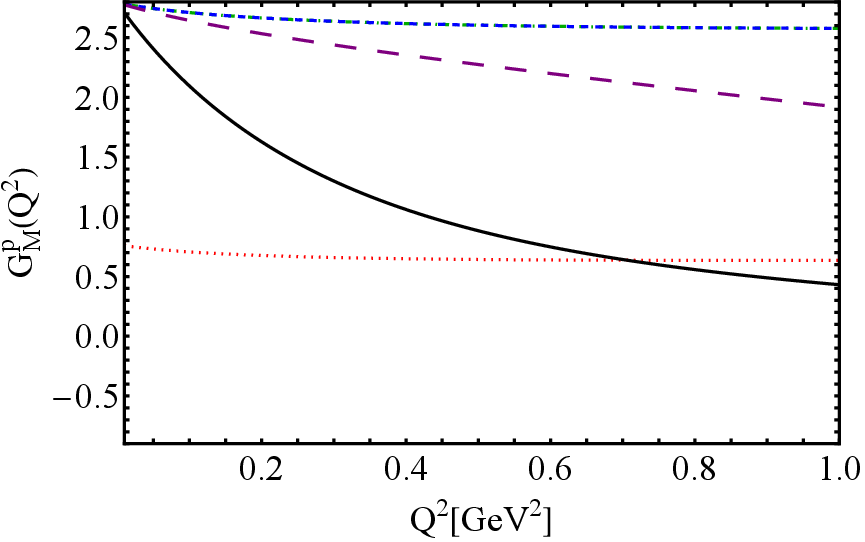}}& \\
\hspace{0.1cm}{\epsfxsize=3.2in\epsfbox{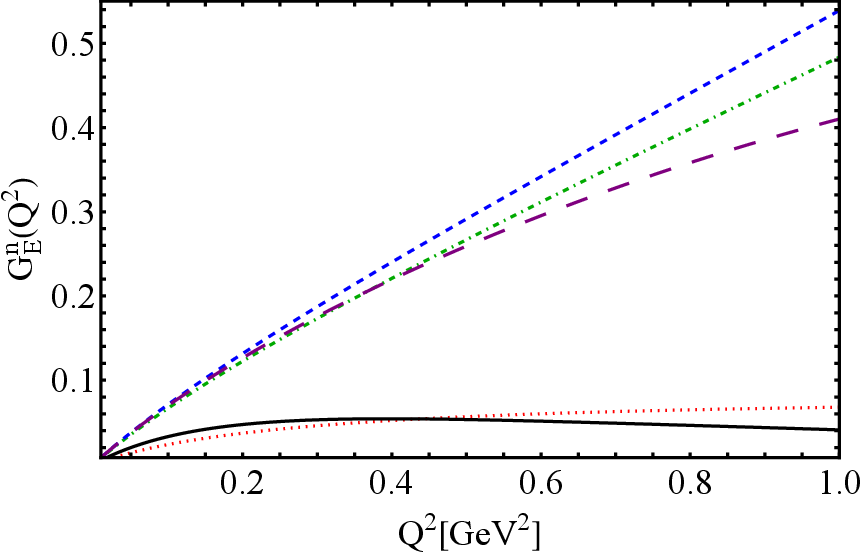}}&
\hspace{0.1cm}{\epsfxsize=3.2in\epsfbox{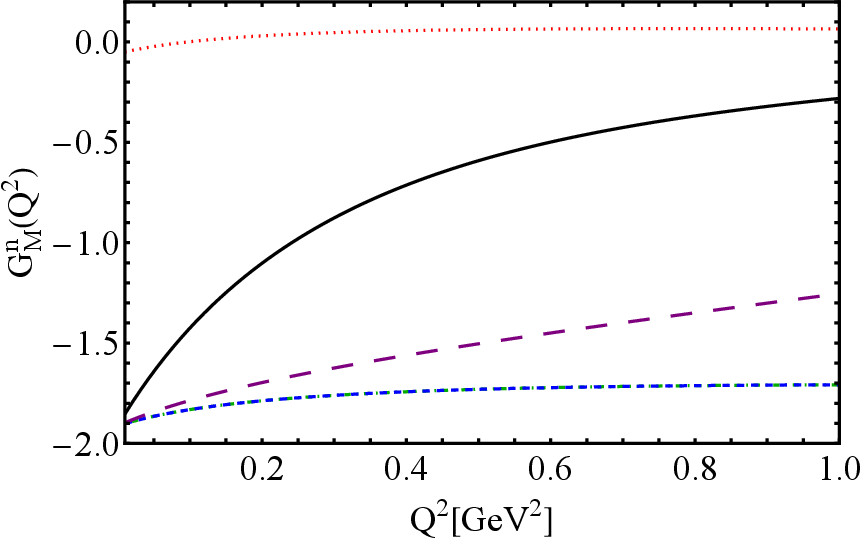}}&
\end{tabular}
\caption{The $Q^2$-dependence of the  nucleon electromagnetic form factors at $\mathcal{O}(p)$(red dotted curve), $\mathcal{O}(p^2)$(green dot-dashed curve), $\mathcal{O}(p^3)$(blue short dashed curve), $\mathcal{O}(p^4)$(purple long dashed curves) and for the result with vector mesons (black solid curve).}
\label{fig:4}
\end{figure*}

In Fig.~\ref{fig:4}, we display the nucleon electromagnetic form factors at different chiral orders in the region $0\leq Q^2\leq 1\ \rm GeV^2$. It is obvious that the $\mathcal{O}(p)$, $\mathcal{O}(p^2)$ and $\mathcal{O}(p^3)$ order pionic contributions exhibit approximately linear behavior in $Q^2$ in this region at all chiral orders. In particular, the $\mathcal{O}(p)$ contributions depend weakly on $Q^2$. These results are consistent with the conclusions from the local chiral perturbation theory~\cite{Fuchs:2003ir,Kubis:2000zd}, and imply that higher-chiral-order and vector-meson contributions are necessary for describing the electromagnetic radii and curvatures.

It is worthwhile to mention that at $\mathcal{O}(p^3)$, the proton and neutron Pauli form factors are equal in magnitude but opposite in sign. Consequently, the magnetic form factors vanish at this order. Nevertheless, the nucleon electric form factors acquire small but  nonzero contributions. The inclusion of the vector-meson contribution, however, significantly modifies the curvature of the form factors due to the vector-meson pole dominance. Consequently, the total electromagnetic form factors are reshaped, and leading to better agreement with experimental data.

In Tab.~\ref{tab3.}, we list the contributions of the different chiral orders to the electromagnetic radii. Note that the pion $\mathcal{O}(p)$, $\mathcal{O}(p^2)$ and $\mathcal{O}(p^3)$ contributions to the proton charge radius are $0.088\rm fm^2 $, $0.110\rm fm^2 $ and $0.068 \ \rm fm^2 $, respectively, yielding a total pionic contribution is $0.266 \ \rm fm^2$. In addition, the  vector-meson contribution is $0.441 \rm fm^2$, which accounts for about $62.4\%$ of the proton charge radius. This result indicates that vector-meson contributions are important component of the  nucleon electromagnetic structure. For the proton magnetic radius, the total pionic contribution is $0.154\rm fm^2$, while the vector meson contribution is $0.57\rm fm^2$, which corresponds to  about $78.6\%$ of the total magnetic radius. These conclusions agree with the results from local chiral perturbation theory \cite{Kubis:2000zd}. In the neutron case, the vector meson contribution partially cancels the negative pionic contributions to the neutron electric charge radius and resulting in  the total value $-0.115\rm \ fm^2$. Conversely, the vector meson contribution is responsible for approximately $73.2\%$ of the total neutron magnetic radius. This confirms again  the non-negligible role of vector mesons in shaping the neutron magnetic structures.                          
\begin{table*}
\caption{Contribution of different chiral orders to the nucleon electromagnetic  radii.}
\begin{center}
\begin{tabular}{c|cccccccccccccccccccc}
  \hline
  \hline
   &&& $\mathcal{O}(p)$&&& $\mathcal{O}(p^2)$ &&& $\mathcal{O}(p^3)$  &&& $\mathcal{O}(p^4)$ &&& vector meson &&& total  \\  \hline
  $\langle r^2\rangle^p_E  \rm ( fm^2 ) $ &&& 0.088 &&& 0.110&&&0.068 &&&0&&& 0.441&&&0.707\\ 
  \hline
  $\langle r^2\rangle^p_M   \rm ( fm^2 ) $ &&& 0.101 &&& $-0.001$&&&0 &&&0.055&&&0.569&&&0.724\\
  \hline
  $\langle r^2\rangle^n_E  \rm ( fm^2 )$  &&&$-0.086$ &&& $-0.094$&&&$-0.010$&&&0&&&0.074&&&$-0.116$\\
  \hline
 $\langle r^2\rangle ^n_M  \rm ( fm^2 )$  &&&0.146 &&& $-0.001$&&&0 &&&0.055&&&0.547&&&0.747 \\
  \hline
  \hline
\end{tabular}
\end{center}
\label{tab3.}
\end{table*}

\section{summary}
\label{sec6}
 In this work, to cure the UV divergences of pion loop integrals, we first introduce nonlocal meson fields and construct the nonlocal chiral Lagrangians for the pion-nucleon and vector meson-nucleon interactions. Within this framework, we then analytically derive the pion one-loop corrections to the nucleon electromagnetic form factors up to $\mathcal{O}(p^4)$ order. Due to the nonlocality, the loop corrections contain a regulator function $\tilde{F_\phi}^2(k)$, which modifies the superficial degree of divergence of the loop integrals. As a result, the loop integrals are UV convergent with a cutoff mass $\Lambda_\phi$. 
 
 Meanwhile, because the chiral low‑energy coupling constants depend on the renormalization scheme, the low-energy coupling constants for the pion-nucleon and vector meson-nucleon interaction are fitted to the $Q^2$-dependence of the nucleon electromagnetic form factors within the nonlocal framework. The least-squares fit analysis yields the vector meson coupling constants and the vector meson cutoff mass as $g_{e\rho}=5.241 $, $g_{e\omega}=25.197$, $g_{m\rho }/g_{e\rho }=6.183$, $ g_{m\omega }/g_{e\omega }=0.055$ and $\Lambda_\rho=1.79\ \rm GeV$. By comparing, we find that the fitted values for the vector-to- tensor coupling constant ratios $\rm g_{m\rho}/g_{e\rho}$ and $\rm g_{m\omega}/g_{e\omega}$ as well as the $\Lambda_\rho$ are consistent with the ranges extracted from nucleon-nucleon scattering, while the vector coupling constants $g_{e\rho}$ and $g_{e\omega}$ are larger than the upper limits of those ranges. Correspondingly, the $\mathcal{O}(p^2)$, $\mathcal{O}(p^3)$ and $\mathcal{O}(p^4)$ chiral low energy coupling constants are fixed as $c_6=6.263$, $c_7=-2.979$, $d_6=-0.247{\rm GeV^{-2}}$, $d_7=-0.092{\rm GeV^{-2}}$, $e_{54}=0.041 {\rm GeV^{-3}}$ and $e_{74}=0.438 {\rm GeV^{-3}}$, respectively. We find that the third and  fourth-order LECs are totally different from those in the local case, whereas the absolute values of the second-order low energy coupling constants are significantly larger than the local values.

With these fitted LECs and cutoff masses, we compute the $Q^2$-dependence of the nucleon electromagnetic form factors in the $0\leq Q^2\leq 1\ \rm GeV^2$ region. The model results show that the nonlocal method greatly improves the $Q^2$-dependence of the nucleon electromagnetic form factors at the large-$Q^2$ region. In particular, the vector meson contributions are essential for describing the electromagnetic properties in the large-$Q^2$ region. Moreover, the model calculation yields the fourth moments of the nucleon electromagnetic form factors as $\langle r^4\rangle^p_E=1.331 \ [\rm fm^4]$, $\langle r^4\rangle^p_M=1.380 \ [\rm fm^4]$, $\langle r^4\rangle^n_E=-0.422 \ [\rm fm^4]$, $\langle r^4\rangle^n_M=1.516 \ [\rm fm^4]$, which are smaller than the central values of the experimental data. We therefore expect that these results for the fourth moments will provide a prediction for future experimental observations.  

 \section*{Acknowledgments}
This work is supported  by the National Natural Science Foundation of China under
Grant No. 12265016.
\appendix
\section{Vertex functions from individual Feynman diagrams}
\label{sec.6}
\subsection{$\mathcal{O}(p)$ contributions  }
\begin{equation}
\Gamma_{a}^\mu(p,q)=\frac{1+\tau^3}{2} \bar u(p') \gamma^\mu   u(p),
\end{equation}
where the isospin index $\tau^3$ takes $\tau^3=\pm1$ for the proton and neutron.
\begin{eqnarray}
 &&\Gamma_b^\mu (p,q)  =\frac{-(3-\tau^3)g^2_A }{8 f ^2}
  \bar u(p')  \int \frac{ d^4k} {{(2\pi)}^4}  (-i \centernot k \gamma ^5)   \frac{i}{{D_N} (p'-k)} \nonumber\\
  &&\gamma^\mu    \frac{i}{{D_N} (p-k)}    (i \centernot k \gamma ^5)  \frac{i}{{D_\pi} (k) }   u(p)\widetilde{F}_{\pi}^2(k),   
\end{eqnarray}
where  ${D_N} (k)= \centernot k-m_N+i \epsilon $, ${D_\pi} (k)=  k ^2-m_\pi^2+i \epsilon $, $\widetilde{F}_{\pi}(k)$ is the Fourier transform of the nonlocal regulator function $\widetilde{F}_{\pi}(k)=\int^4 \frac{d^4a}{(2\pi)^4} F_{\pi}(a)e^{-ik.a}$.
\begin{eqnarray}
 &&\Gamma_c^\mu  (p,q)  =\frac{ g^2_A}{ 2f ^2 }
  \bar u(p') \int \frac{ d^4k} {{(2\pi)}^4}    [-i (\centernot k+\centernot q) \gamma ^5]  \frac{i}{{D_N} (p-k)} \nonumber \\
 && \frac{i }{D_\pi (k+q)}  (2k+q)^\mu  \frac{i }{ D_\pi (k)}  (i \centernot k \gamma ^5)  u(p) \widetilde{F}_{\pi}(k) \widetilde{F}_{\pi}(k+q). \nonumber \\ 
\end{eqnarray}
\begin{eqnarray}
&&\Gamma_{d+e}^\mu  (p,q) =\frac{-\tau^3 g^2_A  }{2 f ^2 } \bar u(p')  \int \frac{ d^4k} {{(2\pi)}^4} [  (-i\centernot k \gamma ^5)  \frac{i }{{D_N} (p'-k)}  \nonumber \\
 &&\frac{i }{ {D_\pi} (k)}  \gamma ^\mu   \gamma ^5 \widetilde{F}_{\pi}(k-q) \widetilde{F}_{\pi}(k)+ \gamma ^\mu   \gamma ^5  \frac{i}{  {D_N} (p-k)}  \nonumber\\
 &&\frac{i }{{D_\pi} (k)}  (-i\centernot k   \gamma ^5)  \widetilde{F}_{\pi}(k+q)\widetilde{F}_{\pi}(k)   ] u(p).   
\end{eqnarray}

\begin{eqnarray}
&&\Gamma_{f+g}^\mu(p,q)   =\tau^3    \frac{    g^2_A }{  2f ^2 } \int^1_0 dt\int \frac{ d^4k} {{(2\pi)}^4}
  \bar u(p')\Big\{  - (-i \centernot k \gamma ^5) \nonumber\\
  &&\frac{i}{D_N(p'-k)}  \frac{i }{ D_\pi(k)}  ( \centernot k   \gamma ^5)   \frac{\partial \widetilde{F}_{\pi}(k-qt)}{\partial k_\mu  }\widetilde{F}_{\pi}(k)+( \centernot k \gamma ^5)  \nonumber \\
  &&  \frac{\partial\widetilde{F}_{\pi}(k+qt)}{\partial k_\mu  } \widetilde{F}_{\pi}(k)    \frac{i}{D_N(p-k)} \frac{i}{ D_\pi(p-k)}   ( -i\centernot k   \gamma ^5) \Big  \}  u(p). \nonumber \\
\end{eqnarray}
\begin{eqnarray}
 &&\Gamma_{h}^\mu (p,q) =\frac{-\tau^3}{4f^2}
  \bar u(p')    \int \frac{ d^4k} {{(2\pi)}^4} \frac{i }{ {D_\pi} (k)} \gamma^\mu      u(p) [\widetilde{F}_{\pi}(k-q)\nonumber \\
  &&+\widetilde{F}_{\pi}(k+q)] \widetilde{F}_{\pi}(k). 
 \end{eqnarray}
 \begin{eqnarray}
&&\Gamma_{i}^\mu(p,q)  =\frac{-i \tau^3}{4f^2} \bar u(p')\int  \frac{ d^4k} {{(2\pi)}^4}
 \frac{i }{{D_\pi} (k+q)}  \frac{i }{{D_\pi} (k) } \nonumber\\
  && (\centernot q+2\centernot k )   (2 k+q)^\mu   u(p) \widetilde{F}_{\pi}(k+q) \widetilde{F}_{\pi}(k). \nonumber\\
\end{eqnarray}
\begin{eqnarray}
&&\Gamma_{j}^\mu(p,q) = \frac{i\tau^3 }{ 4f^2} \bar u(p')\int^1_0 dt \int  \frac{ d^4k}  {{(2\pi)}^4} \frac{i }{ D_\pi (k)}
   2i\centernot k    \Big \{    \frac{\partial \widetilde{F}_{\pi}(k+tq)}{\partial k_\mu } \nonumber\\
  && \widetilde{F}_{\pi}(k)+\frac{\partial \widetilde{F}_{\pi}(k-tq)}{\partial k _\mu }\widetilde{F}_{\pi} (k)  \Big  \}  u(p).
\end{eqnarray}
\subsection{$\mathcal{O}(p^2)$ contributions}
\begin{equation}
\Gamma^\mu_{k}(p,q)=[\frac{(1+\tau^3)}{2}c_6+c_7] \bar  u(p') \frac{i \sigma^{\mu\nu}q^\nu }{2m} u(p).
\end{equation}

\begin{eqnarray}
&&\Gamma^\mu_{l}(p,q) = \frac{1 }{f^2}
  \bar u(p')    \int \frac{ d^4k} {{(2\pi)}^4} \frac{i }{ {D_\pi}(k)  }  [-\tau^3\frac{c_6}{2} \frac{i \sigma^{\nu\mu}q^\nu}{2m}+3(\tau^3
  \nonumber\\
  &&
  +1)\frac{c_2}{2m^2}(p.k+p'.k) k_\mu]    u(p) \widetilde{F}_{\pi}^2(k). \nonumber \\   
\end{eqnarray}

\begin{eqnarray}
&&\Gamma^\mu_{m}(p,q) =\frac{i\tau^3}{f^2}  \bar u(p')
\int \frac{ d^4k} {{(2\pi)}^4} \frac{i }{ {D_\pi}(k+q)}  \frac{i }{ {D_\pi}(k)}
   \biggl\{ \frac{c_2}\nonumber\\
   &&{m^2}\biggr[(p.k)(p.k')+(p'.k)(p'.k')\biggr]+c_4 \biggl[\centernot k,\centernot k'\biggr]\biggr\} (2k+q)^\mu\nonumber \\
   &&\widetilde{F}_{\pi} (k+q)\widetilde{F}_{\pi}(k)  u(p).\nonumber \\
\end{eqnarray}

\begin{eqnarray}
 &&\Gamma^\mu_{n} (p,q) =\frac{  g^2_A [c_6(3-\tau^3)+6c_7] }{32 f ^2 }
  \bar u(p) \int \frac{ d^4k} {{(2\pi)}^4}  (-i \centernot k \gamma ^5)\nonumber\\ &&\frac{i }{ {D_N}(p'-k)} \nonumber (-i2q^\nu \sigma^{\nu\mu})  \frac{i}{ {D_N}(p-k)}    (i \centernot k \gamma ^5)  \frac{i  }{{D_\pi}(k)}   u(p)\hfill  \widetilde{F}_{\pi}^2(k). \nonumber\\
\end{eqnarray}

\subsection{$\mathcal{O}(p^3)$  contribution}
\begin{equation}
\Gamma^\mu_{o}(p,q) = -t(\tau^3 d_6+2 d_7)  u(p') \frac{(p+p')_\mu}{2 m} u(p).
\end{equation}
\subsection{ $\mathcal{O}(p^4)$  contribution}
\begin{equation}
\Gamma^\mu_{p}(p,q)=2m_Nt(\tau^3 e_{54}+2e_{74}) \bar  u(p') \frac{i \sigma^{\mu\nu}q^\nu }{2m} u(p).\nonumber \\
\end{equation}
\subsection{Vector meson contribution}
\begin{eqnarray}
&&\Gamma^\mu_{q}(p,q)=-[\tau^3\frac{q^2 F_\rho g_{e\rho}}{q^2-m_V^2}\widetilde{F}_{\rho}^2(k)+\frac{q^2 F_\omega g_{e\omega}}{(q^2-m_V^2)}\widetilde{F}_{\omega}^2(k)]\nonumber \\
&&\bar u(p') \gamma^\mu   u(p)-[\tau^3\frac{q^2 F_\rho g_{e\rho}}{q^2-m_V^2}\widetilde{F}_{\rho}^2(k)+\frac{q^2 F_\omega g_{e\omega}}{(q^2-m_V^2)}\widetilde{F}_{\omega}^2(k)]\nonumber\\
&&\frac{i \sigma^{\mu\nu}q^\nu }{2m} u(p).\nonumber \\ 
\label{eq:A15}
\end{eqnarray}

%%%%%%%%%%%%%%%%%%%%%%%%%%%%%%%%%%%%%%%%%%%%%%%%%%%%%%%%%%%%%%%%%%%%%%%%

\end{document}